\documentclass[conference]{IEEEtran}
\IEEEoverridecommandlockouts
\usepackage{indentfirst}
\usepackage{subfigure}
\usepackage{graphicx}
\usepackage{mathrsfs}
\usepackage{bm}
\usepackage{amsmath}
\usepackage{amsfonts,amssymb}
\usepackage{color}
\usepackage{algorithm}
\usepackage{algpseudocode}
\usepackage{amsmath}
\usepackage{cite}
\usepackage{multirow}
\usepackage{booktabs}
\usepackage{float}
\usepackage{amsmath}
\usepackage[numbers,sort&compress]{natbib}
\def\BibTeX{{\rm B\kern-.05em{\sc i\kern-.025em b}\kern-.08em
    T\kern-.1667em\lower.7ex\hbox{E}\kern-.125emX}}

\ifCLASSINFOpdf

\else

\fi

\begin{document}

\title{A Content Driven Resource Allocation Scheme for Video Transmission in Vehicular Networks\\
\thanks{This work was supported by the Beijing Natural Science Foundation under Grant No. 4202049 and the National Key R\&D  Program of China under Grant No. 2018YFB1800805.}
}

\author{\IEEEauthorblockN{Jiujiu~Chen\IEEEauthorrefmark{1}, Chunyan~Feng\IEEEauthorrefmark{1}\,
Caili~Guo\IEEEauthorrefmark{2}, Xu~Zhu\IEEEauthorrefmark{1}}
 \IEEEauthorblockA{\IEEEauthorrefmark{1}Beijing Laboratory of Advanced Information Networks \\}
 \IEEEauthorblockA{\IEEEauthorrefmark{2}Beijing Key Laboratory of Network System Architecture and Convergence \\ Beijing University of Posts and Telecommunications, Beijing, 100876, China\\}
 Email: \{chenjiujiu, cyfeng, guocaili, zhuxu\}@bupt.edu.cn}

\maketitle

\begin{abstract}
With the growing computer vision applications, lots of videos are transmitted for content analysis, the way to allocate resources can affect the performance of video content analysis. For this purpose, the traditional resource allocation schemes for video transmission in vehicular networks, such as quality-of-service (QoS) based or quality-of-experience (QoE) based schemes, are no longer optimal anymore.  In this paper, we propose an efficient content driven resource allocation scheme for vehicles  equipped with cameras under bandwidth constraints in order to improve the video content analysis performance. The proposed resource allocation scheme is based on maximizing the quality-of-content (QoC), which is related to the content analysis performance. A QoC based assessment model is first proposed. Then, the resource allocation problem is converted to a solvable convex optimization problem. Finally, simulation results show the better performance of our proposed scheme than the existing schemes like QoE based schemes.
\end{abstract}


\IEEEpeerreviewmaketitle

\section{Introduction}

Recently vehicular networks get wide attentions, and many researches focus on communications in vehicular networks [1]. At the same time, video service becomes a large proportion of wireless traffic in vehicular networks and takes up a lot bandwidth, which makes wireless resource allocation in vehicular networks very important [2]. Traditional resource allocation schemes focus on quality-of-service (QoS) based or quality-of-experience (QoE) based methods.
For the QoS based schemes, resource allocation is optimized by adjusting jointly those network parameters like delay, packet loss, throughput, jitter and so on [3,4]. More followed researches focusing on user experience. For the QoE based schemes, the satisfaction level of users is improved by jointly considering the user perception and experience of videos and the network parameters [5,6].

However, those studies are based on the perspectives of network efficiency and user experience, and do not consider the quality-of-content (QoC) of videos, which is related to content information and content analysis accuracy. Nowadays, many computer vision applications such as autonomous driving focus on visual content understanding and analysis tasks [7,8], where the traditional resource allocation schemes for video transmission is no longer optimal anymore [9]. Hence, under the condition of resource limitation, how to reasonably allocate resources, such as bandwidth, to transfer content information correctly and meet the accuracy requirement of video content analysis tasks, is an urgent problem to be solved.

In this paper, we contribute to design a content driven resource allocation schemes for video transmission in order to get better video content analysis performance. To achieve this purpose, this paper makes the following contributions:

1) We propose a quality assessment model of the video content, named QoC model, which is used as a resource allocation guideline for video transmission from vehicles to the edge server. Based on the basic video content analysis task, taking object detection as an example, the model reveals the relationship between detection accuracy and data rate.

2) We design a QoC based resource allocation scheme in vehicular networks  to get better
video content analysis performance. The resource allocation problem is converted to a convex optimization problem, which can be solved by a standard solver.

The rest of this paper is organized as follows. In Section II, we introduce the system model and video transmission architecture. Section III proposes a QoC model for system performance description and develops a resource allocation scheme based on the QoC formulation. Simulation results are shown in Section IV, followed by the conclusion remarks in Section V.

\section{System Model and Transmission Architecture}
\subsection{System Model}
\begin{figure}[!t]
	\centering
	\includegraphics[width=3.3in]{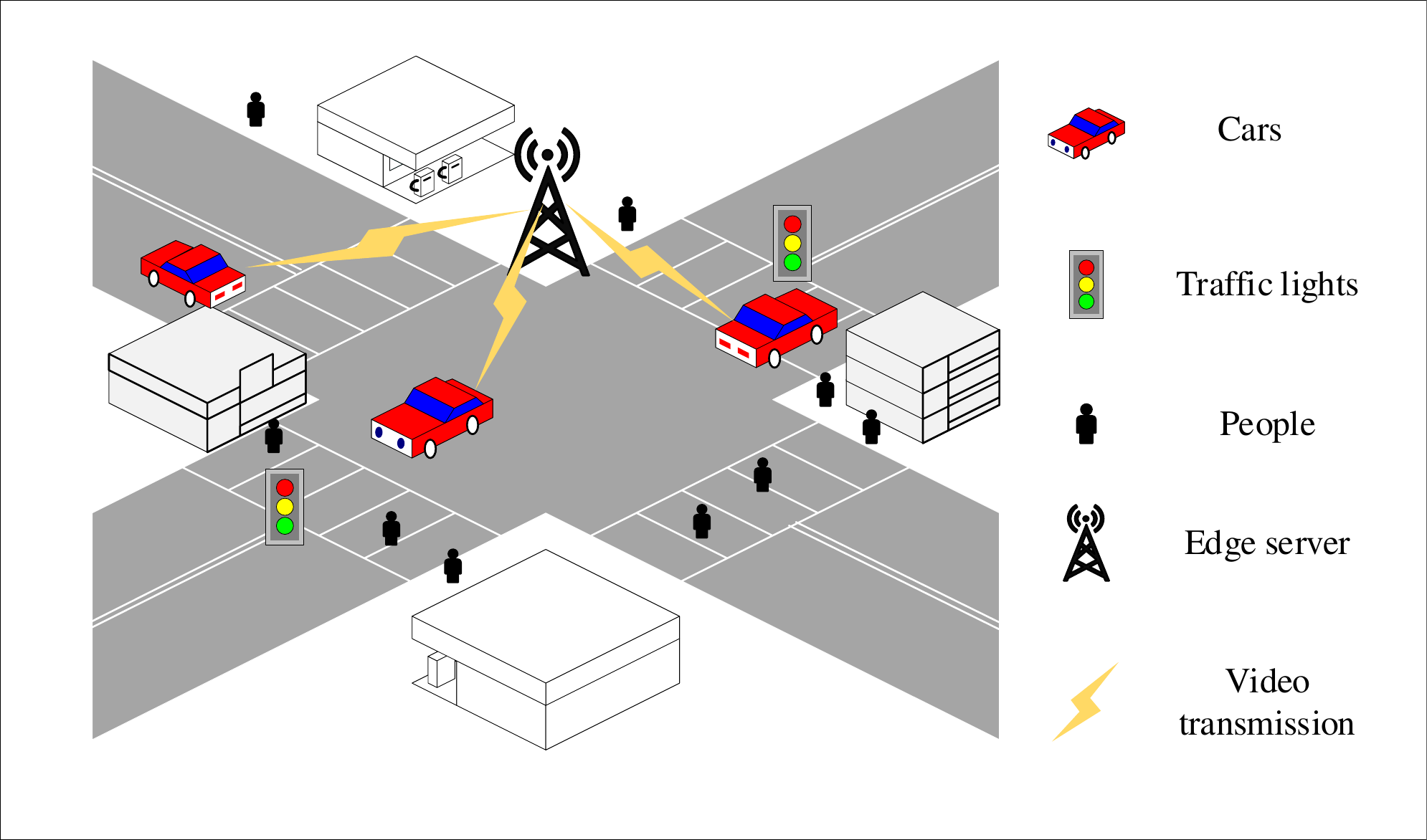}
	\caption{Scenario and system model.}
	\label{System model}
\end{figure}

As shown in Fig. {\ref{System model}}, a vehicular network consists of an edge server and multiple vehicles equipped with cameras. Vehicles are randomly distributed in the server's coverage area and move on different roads, which determine the different collected video content, such as different vehicle and pedestrian densities.Vehicles on each road can encode and transmit the collected videos to the edge server.  The basic video content analysis tasks, such as object detection, are completed at vehicle terminals. Vehicles upload the encoded video, along with the detection results, to the edge server, which allocates resources, such as bandwidth, to vehicles according to the detection results. The captured video clips on vehicles are encoded with the high efficiency video coding (HEVC) [10] with different encoding data rates, based on the detection results, under resource constraint. Therefore, the overall content quality of videos is related to the resource allocation. In this paper, considering that vehicles have certain computational power and energy supply, we assume the video pre-analysis tasks, object detection in this paper, are accomplished at vehicles and video storage or other video content analysis tasks are carried out at the edge server, which can achieve faster analysis results.

\subsection{Transmission Architecture}
\begin{figure}[!t]
	\centering
	\includegraphics[width=3.15in]{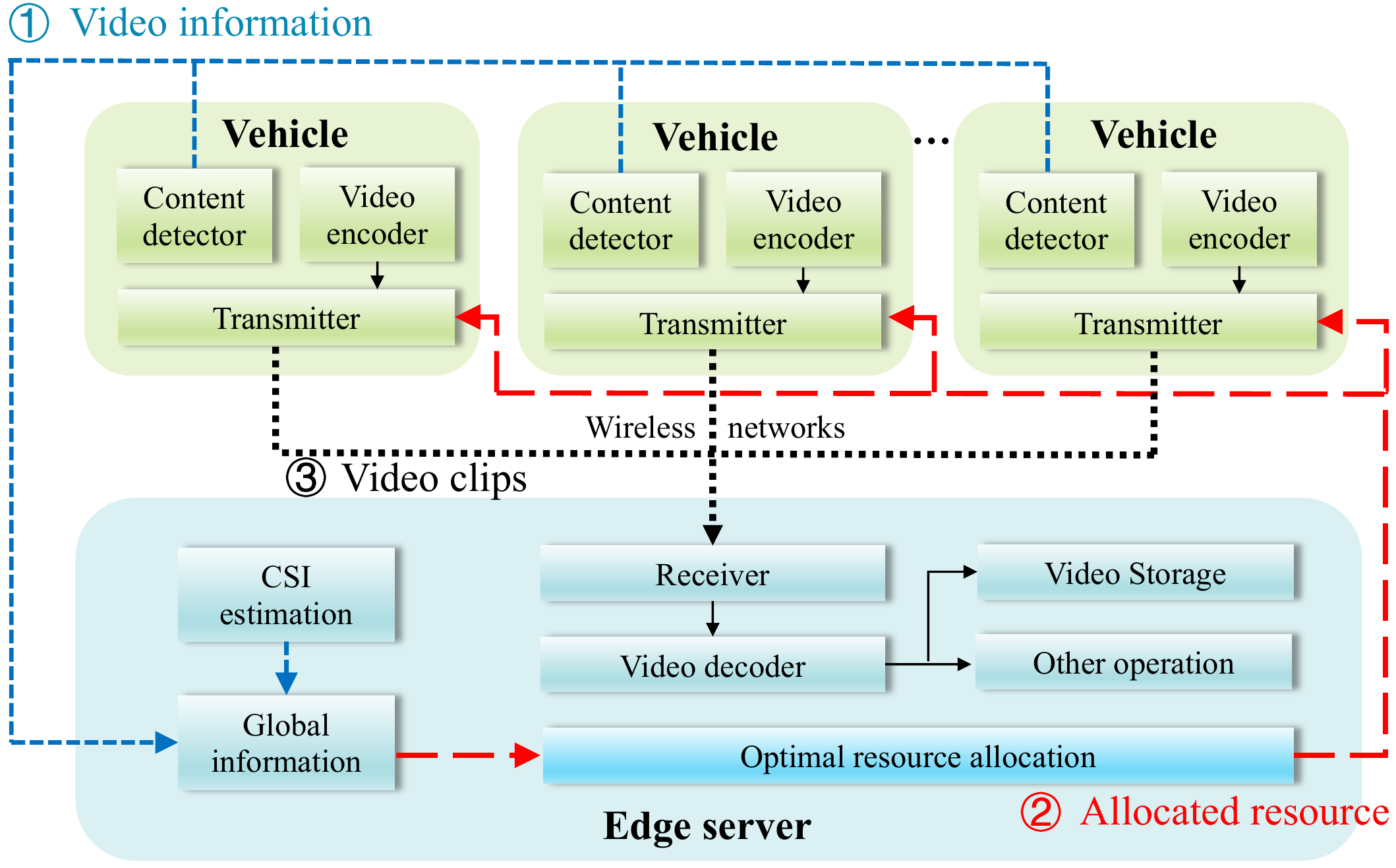}
	\caption{System transmission architecture.}
	\label{transmission structure}
\end{figure}

The system transmission architecture is shown as Fig. {\ref{transmission structure}}. Within the coverage of the edge server, the system completes the transmission of the video collected by the vehicle, assuming that the total duration from the vehicle initiating a video transmission request to the next request is $T$.

In the first step, the vehicles carry out video preprocessing, such as object detection, to obtain the video content information and upload it to the edge server, while the edge server obtains the channel state information (CSI)[11,12]. The first processing delay is $t_1$. In the second step, the edge server obtains global information (including all the video information and CSI), calculates the optimal resource allocation results based on the QoC model, and feeds back to each vehicle end. The second processing delay is $t_2$. Hence, before video clips transmission, the total processing delay is $\Delta t = t_1 + t_2$. Finally, according to the allocated resources, the vehicles encode the video clips and transmits them to the edge server through wireless channels, followed the wireless video transmission protocol. The edge server decodes the received video and stores it or uses it for other content analysis tasks.

\section{QoC Model and Resource Allocation Scheme}
\subsection{Quality Assessment Model of Content}
Based on the system model and transmission architecture, the key to resource allocation is to build the QoC model first. Next, we explore the relationship between content quality and data rate, and described it as a mathematical model. In this paper, quality of video content can be approximately measured by the accuracy of the object detection system. Without loss of generality, we adopt HEVC scheme as the video encoder, which has a wide range of applications. Considering the speed and accuracy of detection and applicability, we adopt the faster region-based convolutional neural networks (Faster R-CNN) [13] as the object detection system in this paper. The object categories mainly include people, cars and traffic lights, which are vital objects in traffic scene. It is worth mentioning that other detection systems or recognition tasks can also be applied or implemented for our proposed scheme.

\begin{figure}[H]
    \centering
        \subfigure[QP=15]{ \includegraphics[width=0.32\hsize]{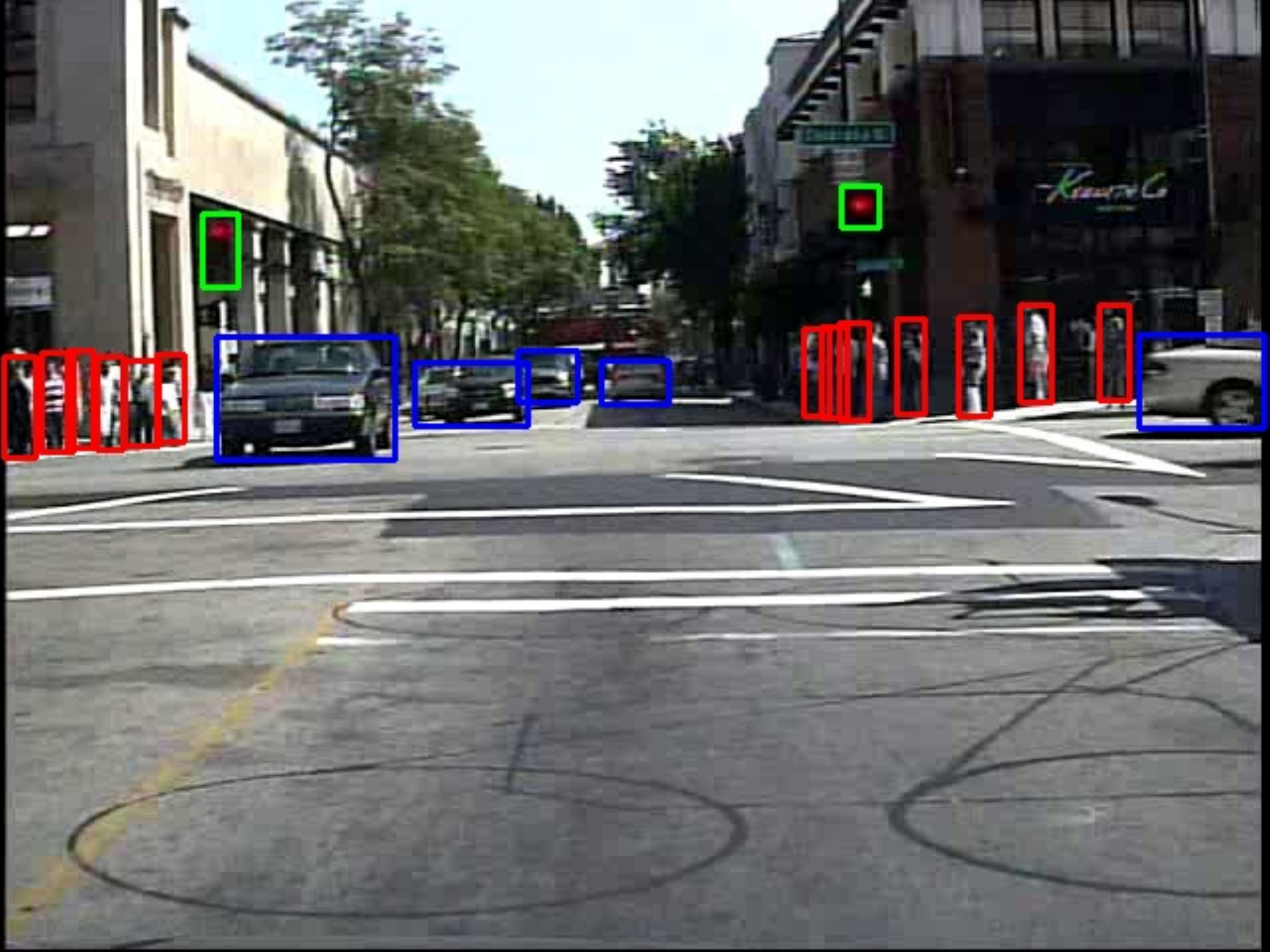}}
        \subfigure[QP=30]{\includegraphics[width=0.32\hsize]{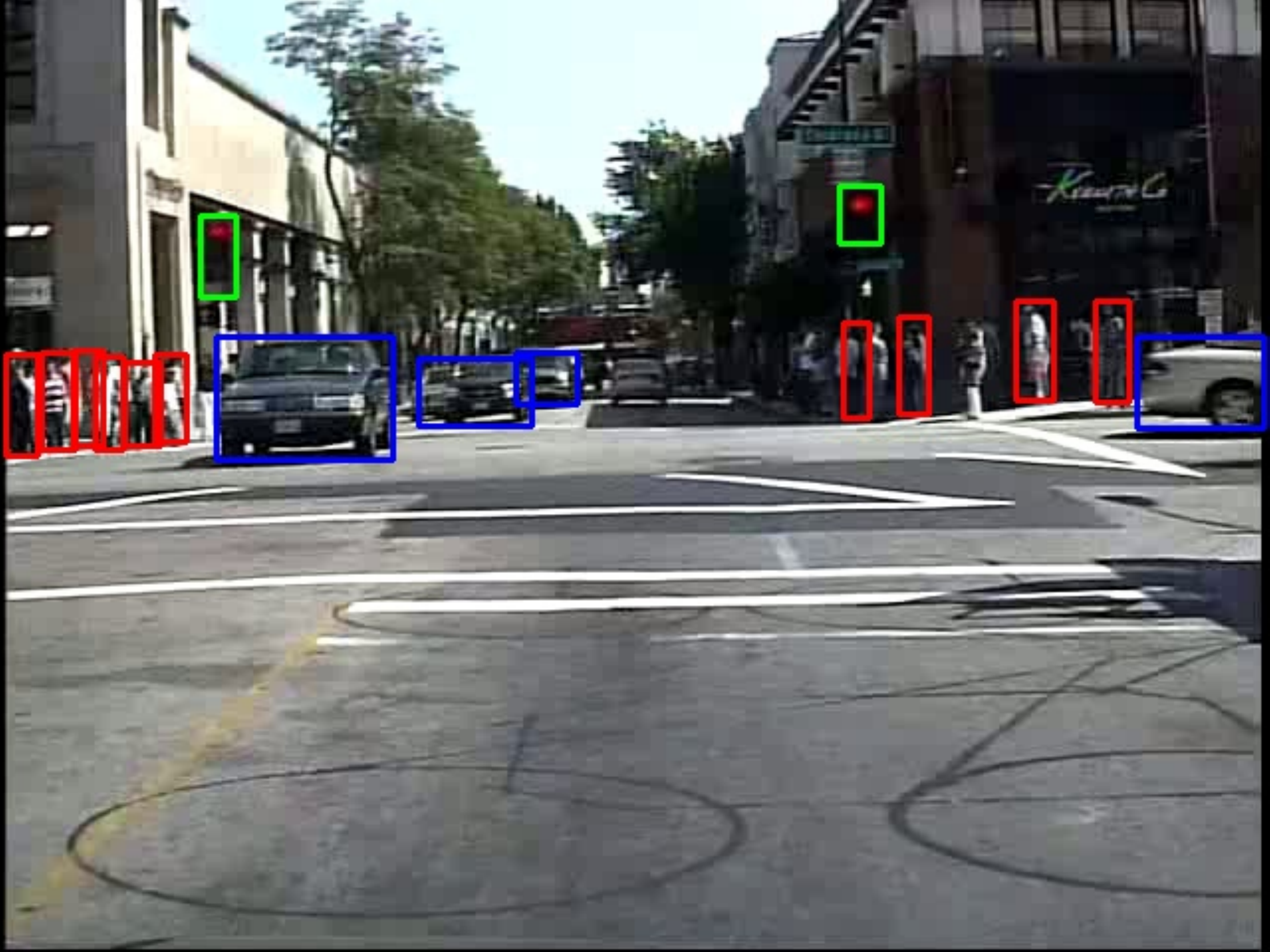}}
        \subfigure[QP=45]{\includegraphics[width=0.32\hsize]{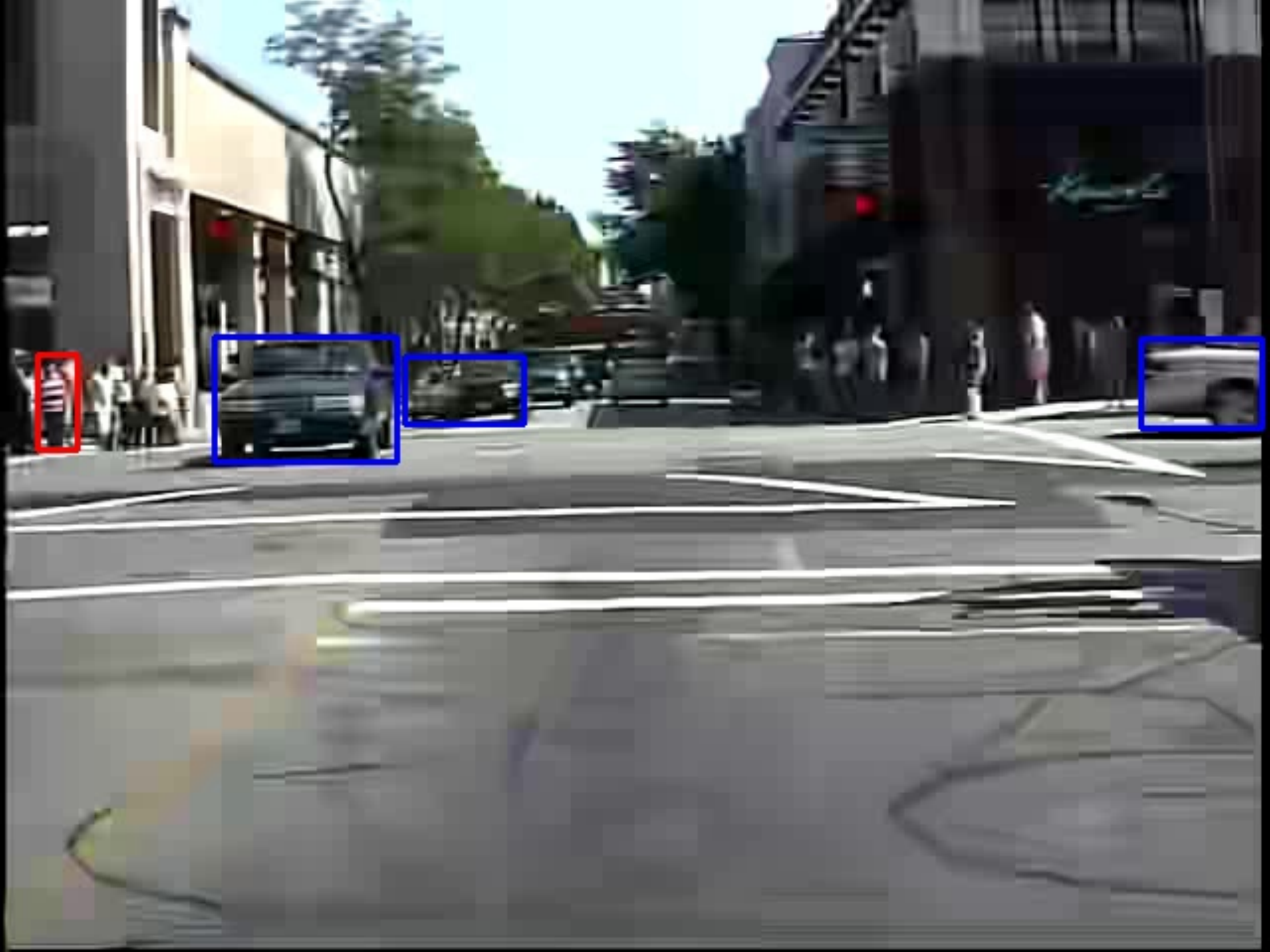}}
           \caption{The qualitatively detection results with different values of QP.}
    \label{The detection results}
\end{figure}

According to [14], the human detection accuracy can be improved by adaptively adjusting the quantization parameters (QP) of video encoder. Figure {\ref{The detection results}} qualitatively shows the detection results on the compressed videos with different values of QP on Caltech Dataset [15]. Red, blue, and green boxes denote people, cars, and traffic lights detection regions, respectively. Different values of QP lead to different compression ratios of video clip, and different video qualities lead to different detection accuracies. Hence, we first find the assessment model of relationship between QP and detection accuracy.

\begin{figure}[!t]
	\centering
	\includegraphics[width=3.5in]{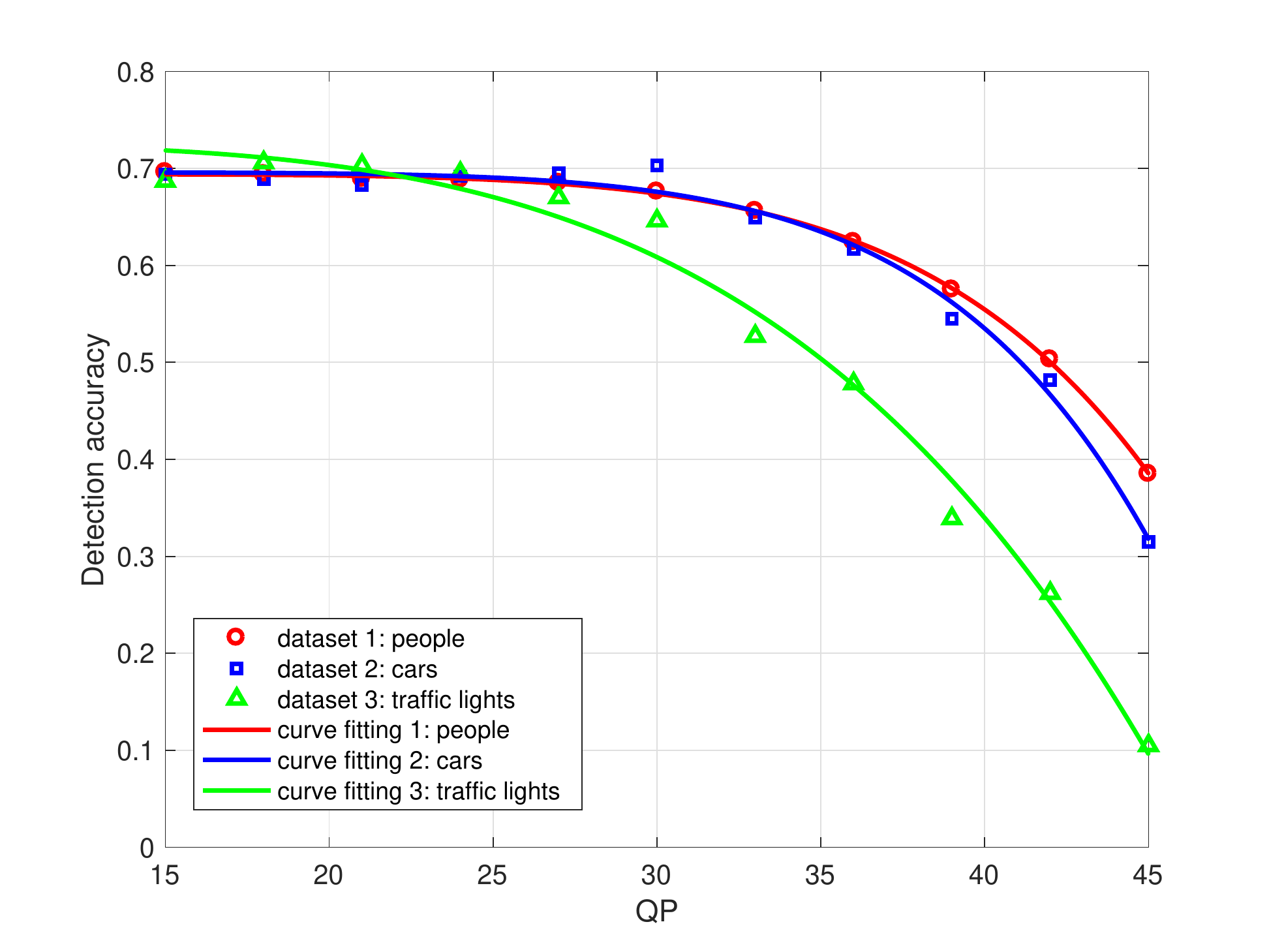}
	\caption{Detection accuracy with different values of QP. }
	\label{detection_accuracy}
\end{figure}

Experiments on Caltech Dataset reveal the relationship between QP and detection accuracy, as shown in Fig. {\ref{detection_accuracy}}, where the markers denote experiment data and solid lines denote fitting curves. According to the results, the higher the value of QP, the lower the detection accuracy. But, the limit value of detection accuracy depends on the detection system itself.

By curve fitting, the different kinds of object detection accuracy can be modeled as a function of QP and can be expressed approximatively as
\begin{equation}
{P_n}\left( Q_m \right) = {\alpha _n} \cdot {Q^{_{{\beta _n}}}} + {\gamma _n},
\label{eq1}
\end{equation}
where $m$ and $n$ denotes the number of videos captured by vehicles (each vehicle collects a video) and the number of detection object categories, respectively. $P_n$ and $Q_m$ denote detection accuracy and value of QP, respectively. $\alpha_n$, $\beta_n$, and $\gamma_n$ are model parameters, which are related to video compression rate and detection accuracy. The root-mean-square error (RMSE) of the curve fittings are 0.208\%, 1.403\%, and 2.603\%, respectively.

\begin{figure}[!t]
	\centering
	\includegraphics[width=3.5in]{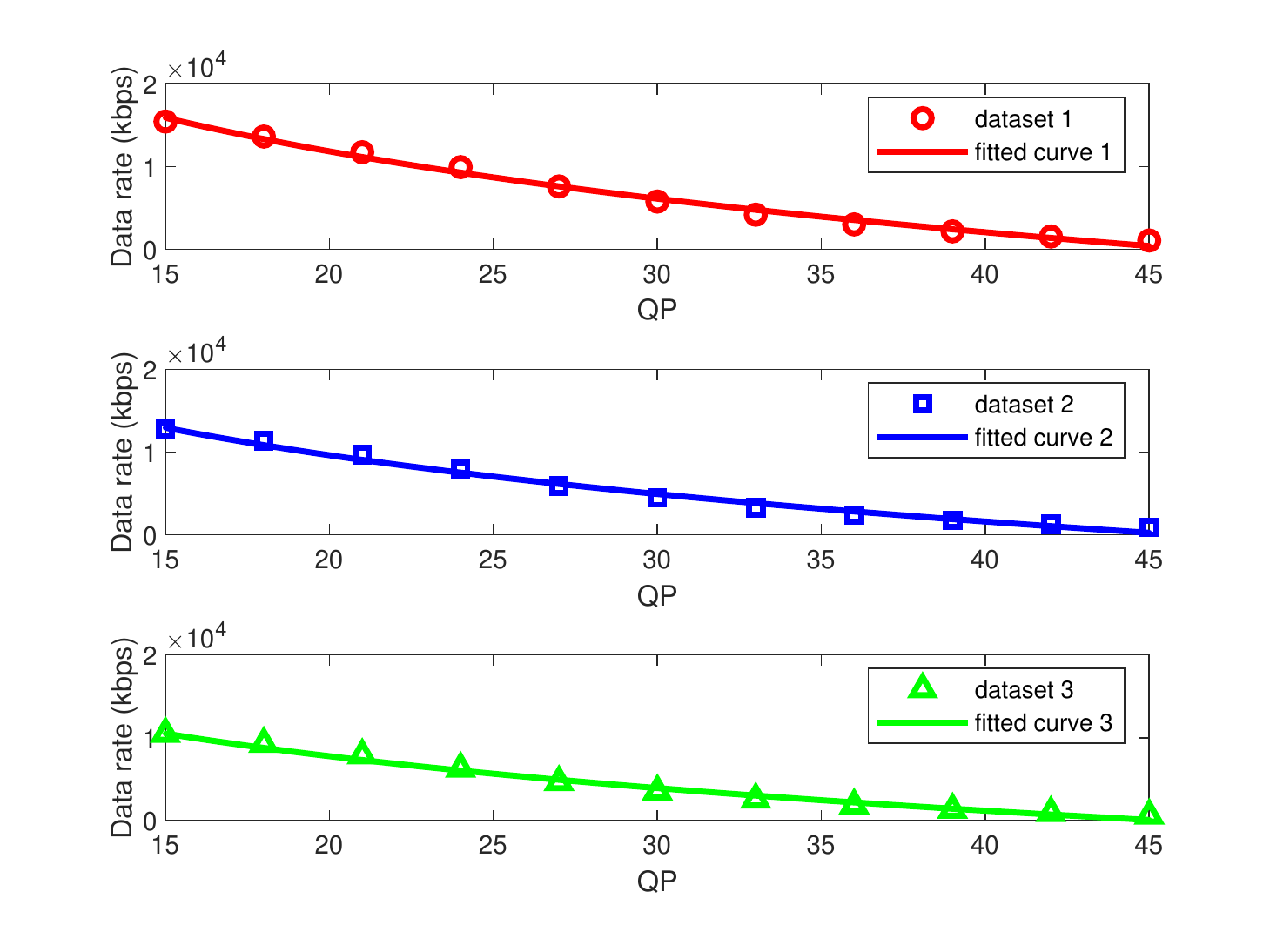}
	\caption{Data rate with different values of QP.}
	\label{datarate}
\end{figure}

As mentioned above, different values of QP determine different compression ratios of video clips, and the sizes of video data rates are related to the compression ratios. Therefore, we then find the relationship between QP and video data rate, as shown in Fig.  {\ref{datarate}}. Obviously, for all video clips from Caltech Dataset, the higher the value of QP, the lower the data rate. By curve fitting, the relationship between QP and data rate can be expressed approximatively as
\begin{equation}
Q_m = {a_m} \cdot {\rm{exp}} ({b_m} \cdot {R_m}),
\label{eq2}
\end{equation}
where $R_m$ denotes data rate. $a_m$ and $b_m$ are model parameters, which are related to video compression rate and video data rate.The RMSE of the curve fittings are 0.198\%, 0.295\%, and 0.405\%, respectively.

Considering different video clips contain different amounts of content information, according to Eq. (\ref{eq1}) and Eq. (\ref{eq2}), the quality assessment model of content for each video can be finally expressed as
\begin{equation}
\begin{aligned}
F_{m}^{\rm{QoC}} &=\sum_{n=1}^{N} I_{m, n} P_{n}\left(Q_{m}\left(R_{m}\right)\right) \\
&=\sum_{n=1}^{N} I_{m, n}\left(\alpha_{n}\left(a_{m} \exp \left(b_{m} R_{m}\right)\right)^{\beta_{n}}+\gamma_{n}\right),
\end{aligned}
\label{eq3}
\end{equation}
where $F_{m}^{\rm{QoC}} $ is the objective function of QoC model, $I_{m, n}$ denotes the amount of content information of different video (i. e. the object density of each video) and $N$ is the total number of object categories. In the paper, $I_{m, n} P_{n} $ represents the mount of correct transmitted content information.

From Eq. (\ref{eq3}), quality of transmitted video content is related to detection accuracy and data rate. The video data rate is subject to transmission rate. So, improving the overall quality of video content  lies in the efficient allocation of resources. In this paper, we mainly focus on bandwidth resource.

\subsection{QoC based Resource Allocation Scheme}
According to the above analysis, it is vital to design an efficient bandwidth allocation scheme to maximize the overall quality of video content at the edge server with a bandwidth constraint since wireless video transmission is bandwidth consuming and the overall bandwidth is limited.

Firstly, the data rate of the transmitted video is constrained by the transmission rate for each vehicle, ${R_m} \le R^T_m(t)$. Considering the time-varying characteristics of the channel state information, the transmission rate for each vehicle at slot $t$ can be formulated as
\begin{equation}
R_{m}^{T}(t)=B_{m} \log _{2}\left(1+\frac{S_{m} h_{m}(t)}{\sigma^{2}_m}\right),
\label{eq4}
\end{equation}
where $B_m$ is the bandwidth allocated for each vehicle, $S_m$, ${h_m}$ and $\sigma^2_m$ denote transmit power, channel gain and noise power, respectively.

To simplify the analysis, we assume that near perfect CSI is available at the transmitters. The channel gain ${h_m}$ modeled as independent random variables, account for both large-scale fading ${h^L_m}$ (contains path loss ${h_{\rm{pl}}}$ and shadowing $h_{\rm{sd}}$) and small-scale fading effects ${h^S_m}$. Since the large-scale fading of channels is typically determined by vehicle locations, which do not change too much during transmission slots [11] . Here,  the path loss is modeled as ${h_{\rm{pl}} = 148.1+37.6{\rm{log}}_{10}(d_m)(\rm{dB})}$, where $d_m$ (in km) is the distance between the $m$th vehicle and the edge server.  Shadowing is modeled using a log-normal distribution, with a standard deviation of 8 dB and zero mean [11]. However, the small-scale fading components might change. Here, the small-scale fading coefficients, denoted by follow the Rayleigh distribution with unit variance and zero mean. Considering the dynamic nature of the small-scale fading, we model the time-varying Raleigh coefficients as independent first-order autoregressive processes [12], given by
\begin{equation}
h_{m}^{S}(t)=\rho_{m}\left(t_{e}\right) h_{m}^{S}\left(t-t_{e}\right)+e_{h},
\label{eq5}
\end{equation}
where $t_{e}$ is the time interval, $e_{h}$ is the process noise sequence which is drawn from a $\mathcal{CN}(0, 1-\rho^2_{m}\left(t_{e}\right))$ distribution, $\rho_{m}\left(t_{e}\right)$ is the channel autocorrelation function and $\rho_{m}\left(t_{e}\right) = J_0(2{\pi}{v_m}{t_e}/{f_c})$, where $J_0(\cdot)$ is the zero-order Bessel function of the first kind, ${f_c}$ is the band mid-frequency and $v_m$ is the velocity of the $m$th vehicle.

Therefore, the goal of our proposed scheme is to optimally allocate the bandwidth resource of each vehicle under constraints so that the overall accuracy of correct detection is maximized. The optimization problem can be express as
\begin{equation}
\begin{array}{l}
\mathrm{P} 1: \max\limits_{B_{m}} \frac{1}{M} \sum\limits_{m=1}^{M} \sum\limits_{n=1}^{N} \delta_{n} I_{m, n} P_{n}\left(Q_{m}\left(R_{m}\right)\right)\\
\rm{s.t.} \\
\quad   \quad  C 1: R_{m} \leq R_{m}^{T}(t),\quad\forall m\\
\quad   \quad  C 2: \sum\limits_{m=1}^{M} B_{m} \leq B_{\rm {total }}\\
\quad   \quad  C 3: B_{m} \geq B_{\min },\quad\forall m\\
\quad   \quad  C 4: P_{n} \geq P_{\min },\quad\forall n\\
\quad   \quad  C 5: \Delta t \leq t \leq T ,\\
\end{array}
\label{eq6}
\end{equation}
where $M$ is the total number of vehicles, and $N$ is the total number of object categories. $\delta_n$ denotes the weight of different categories of detected objects, which means that different categories of detection object (such as people, cars, and traffic lights) are of different importance in content analysis tasks. $B_m$ is bandwidth allocated for each vehicle. The first constraint in Eq. (\ref{eq6}) is based on information transmission theory. $B_{\rm{total}}$ is the total bandwidth, $B_{\rm{min}}$ is the threshold of bandwidth, and $P_{\rm{min}}$ is the threshold of detection  accuracy.

Firstly, according to Eq.(\ref{eq6}), when $t\in(\Delta t, T)$, the CSI is time varying, and the problem is not convex. But, in a time interval $t_e$, the channel state can be considered as stable state, so the problem can be split into $L$ subproblems, where $L=(T-\Delta t)/t_e$. Secondly, due to $\alpha_n<0$, $\beta_n>1$ and $b_m<0$, then $ {\partial P_{n}}\cdot{\partial Q_{m}}>0$ ($\partial$ is derivation operation) and the function $P_n\left( {Q_m\left( {{R_m}} \right)} \right)$ is monotonically increasing in our domain, which takes the maximum when $R_m=R^T_m(t)$. Finally, in the first transmission time interval, for example, the resource allocation problem P1 can be converted to
\begin{equation}
\begin{array}{l}
\mathrm{P} 2: \max\limits_{B_{m}} \frac{1}{M} \sum\limits_{m=1}^{M} \sum\limits_{n=1}^{N} \delta_{n} I_{m, n} P_{n}\left(Q_{m}\left(R^{T}_{m}\left(B_m,t\right)\right)\right)\\
\rm{s.t.} \\
\quad   \quad  C 1: \sum\limits_{m=1}^{M} B_{m} \leq B_{\rm {total }},\quad\forall m\\
\quad   \quad  C 2: B_{m} \geq B_{\min } ,\quad\forall m\\
\quad   \quad  C 3: P_{n} \geq P_{\min } ,\quad\forall n\\
\quad   \quad  C 4: \Delta t \leq t \leq \Delta t+t_e .\\
\end{array}
\label{eq7}
\end{equation}

Combining Eq. (\ref{eq1}) to (\ref{eq4}), according to the rules of the composite functions, the problem P2 is a convex optimization problem, the prove of it is shown in Appendix A,  which can be solved by a standard solver, such as Matlab-based tool for convex optimization (CVX). In our proposed scheme, the allocated results are updated in every transmitted time interval.

\section{Simulation Results}
\begin{figure}[h!]
    \centering
        \subfigure[Frame of Video 1]{ \includegraphics[width=0.32\hsize]{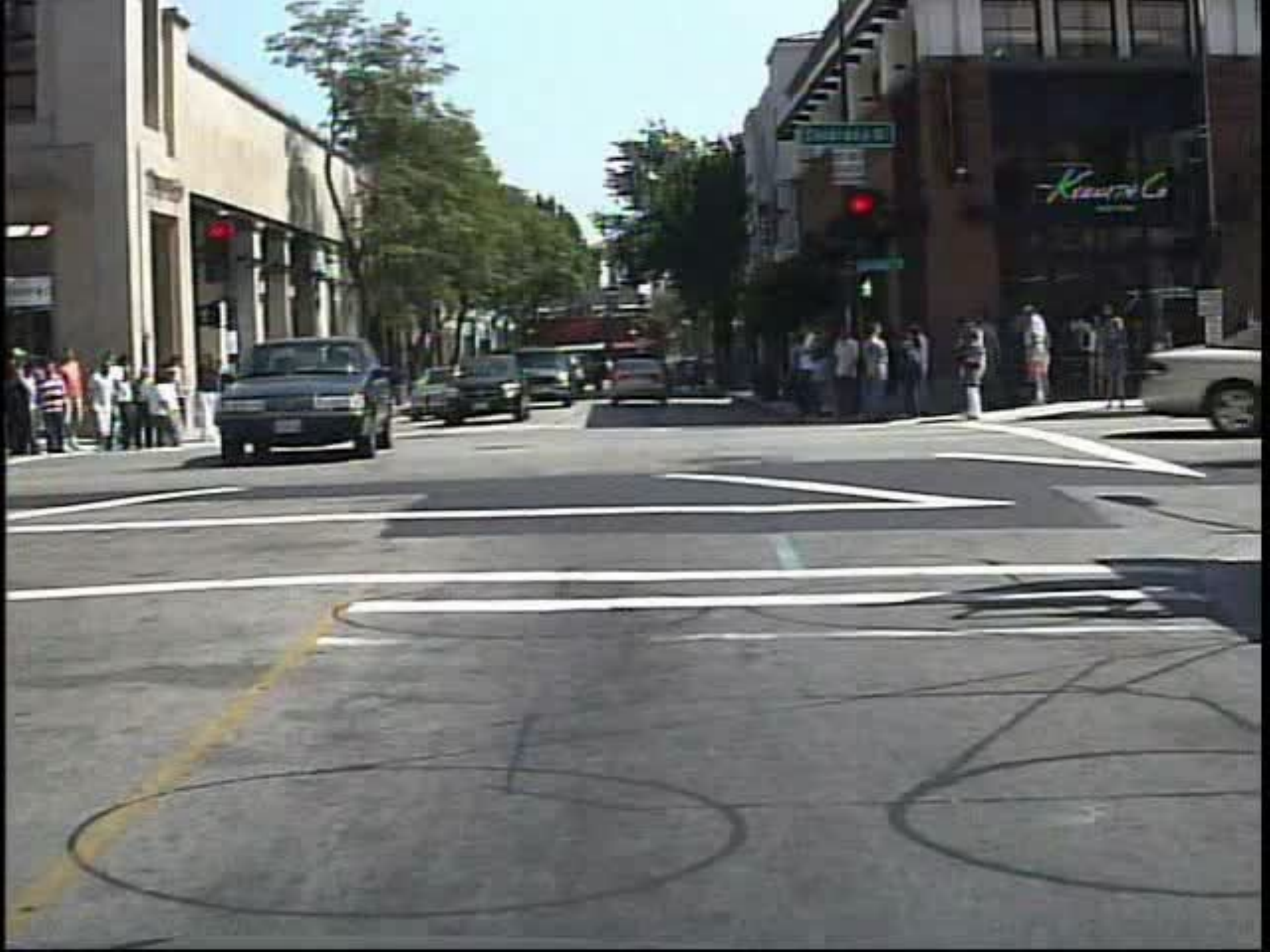}}
        \subfigure[Frame of Video 2]{\includegraphics[width=0.32\hsize]{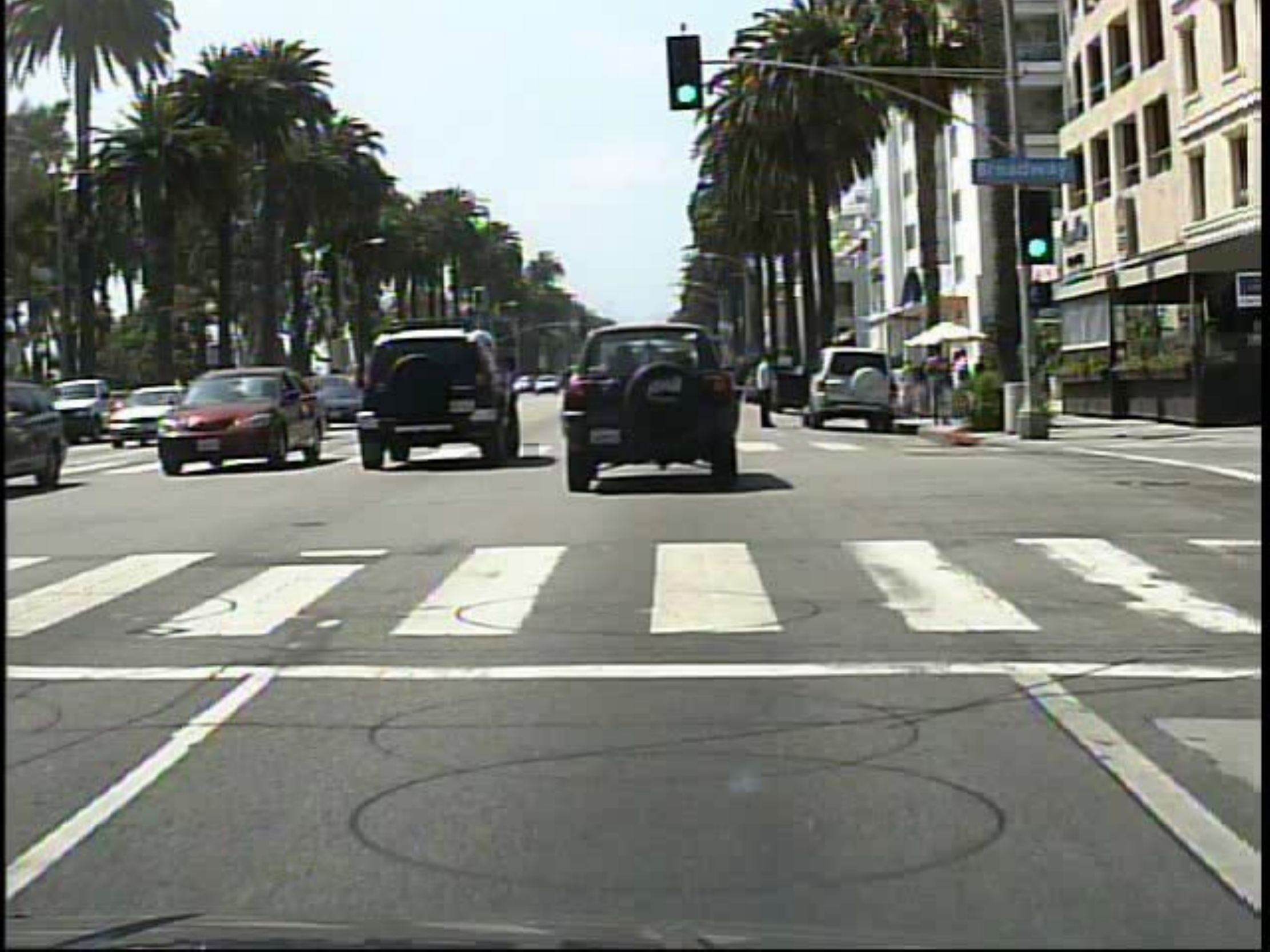}}
        \subfigure[Frame of Video 3]{\includegraphics[width=0.32\hsize]{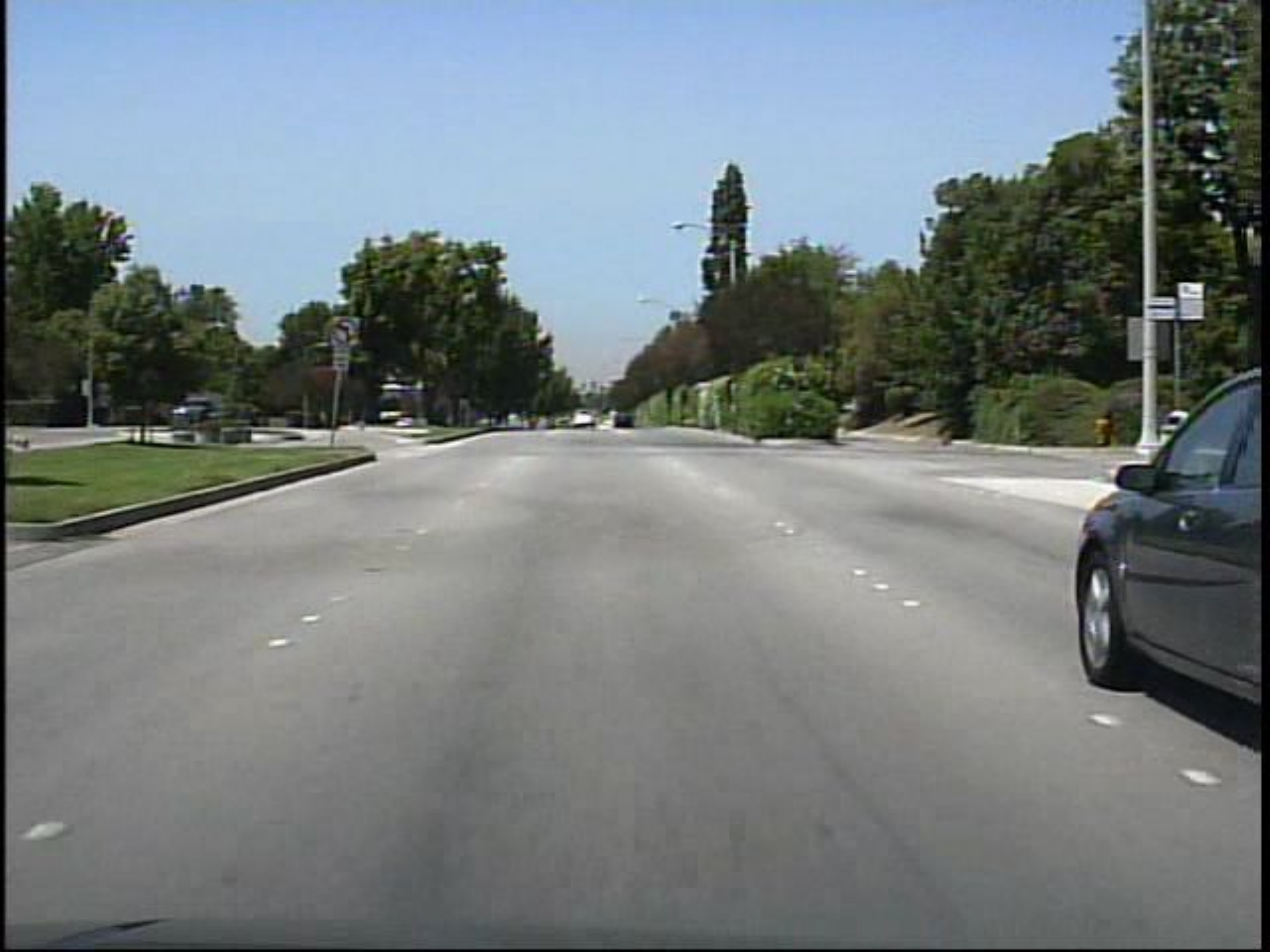}}
           \caption{Sample frames, and the $m$th video is collected by the $m$th vehicle.}
    \label{The sample frames of videos}
\end{figure}

In the simulation experiments, video clips from Caltech Dataset [15] are used for wireless transmission. The video detector uses Faster-RCNN, and the detection environment is Windowns 10 + Compute Unified Device Architecture (CUDA)10.0 + Pytorch 0.4.0. The video encoder uses HEVC,  frame rate  is 30 fps and the group of pictures (GoP) size is 4 (1 I frame and 3 P frames), and 300 frames are tested for each collected video. Sample frames are shown in Fig. {\ref{The sample frames of videos}}.

The system simulation parameters (based on 3GPP TR. 36. 885) and model parameters are given in Table \ref{system parameters} and Table \ref{model parameters} respectively, and Table \ref{video information} shows the information of video clips. The initial relative position of the vehicle to the edge server is randomly generated. Considering the channel characteristics in a vehicular network, 1000 independent Monte Carlo channel simulations are implemented and take the mean to eradicate any discrepancy.

\begin{table}
\centering
\caption{System Simulation Parameters.}
\begin{tabular}{lc}
\toprule
\textbf{ Parameters} & \textbf{Value} \\
\midrule
The edge server coverage & 1.2 km \\
Number of vehicles with cameras $M$ & 3\\
Number of detection  categories $N$ & 3\\
The weights of detection categories $\delta_n, \forall n$ & 1\\
Range of vehicle speed $v_m, \forall m$ & 0$\sim$60 km/h\\
Range of distance $d_m, \forall m$ & 0.1$\sim$1.1 km\\
Transmitted power $S_m, \forall m$ & 23 dBm\\
Noise power spectral density$\sigma^2_m, \forall m$ & -174 dBm/Hz\\
Band mid-frequency $f_c$ & 2GHz\\
The processing delay $\Delta t$ & 200ms\\
The  time interval of small-scale fading $t_e$ & 50ms\\
The total bandwidth $B_{\rm{total}}$ & 2$\sim$20MHZ\\
The minimum bandwidth limit $B_{\rm{min}}$ & $0.1 B_{\rm{total}}$\\
The minimum detection accuracy $P_{\rm{min}}$ & 0.3\\
\bottomrule
\end{tabular}
\label{system parameters}
\end{table}

\begin{table}
\centering
\caption{The Settings of Model Parameters.}
\begin{tabular}{cccccc}
\toprule
$\bm{m}$ \textbf{or} $\bm{n}$ & $\bm {\alpha_n}$ & $\bm{\beta_n}$ & $\bm{\gamma_n}$ & $\bm{a_m}$ & $\bm{b_m}$ \\
\midrule
1 & -2.214e-12 & 6.741 & 0.6940 & 46.27 & -7.086e-5  \\
2 & -3.820e-13 & 7.256 & 0.6958 & 45.96 & -8.648e-5  \\
3 & -8.405e-8 & 4.158 & 0.7250 & 45.22 & -1.052e-4  \\
\bottomrule
\end{tabular}
\label{model parameters}
\end{table}

\begin{table*}
\centering
\caption{The information of video clips.}
\begin{tabular}{cccccc}
\toprule
\textbf{Video} & \textbf{Dataset} & \textbf{Resolution} & \textbf{Pedestrian density} & \textbf{Vehicle density} & \textbf{Number of objects per frame (person/car/traffic light)}\\
\midrule
1 & Caltech set03v008 & 640*480 & High & Low & 11.1 / 1.6 / 1.5\\
2 & Caltech set01v000 & 640*480 & Low & High & 1.0 / 8.5 / 0.3\\
3 & Caltech set10v004 & 640*480 & Low & Low & 0.0 / 1.9 / 0.0\\
\bottomrule
\end{tabular}
\label{video information}
\end{table*}

We compare our proposed QoC based allocation scheme with two other schemes: the detection accuracy (DA) driven scheme in [9], proposed to maximize the mean accuracy, and the QoE based scheme in [5], proposed to maximize the QoE-based mean-opinion-score (MOS) value. The differences and comparison of each scheme are shown in Table \ref{scheme comparison}.

\begin{table}
\centering
\caption{The comparison of each scheme.}
\begin{tabular}{cccc}
\toprule
 & \textbf{QoC-based}  & \textbf{DA-based} & \textbf{QoE-based} \\
\midrule
\textbf{Scheme source}& Proposed & [9] & [5]  \\
\textbf{Guidance Model}& Eq. (3)  & Eq. (1) in [9]  & Eq. (5) in [5] \\
\textbf{Solver}& CVX & CVX & CVX\\
\textbf{Video content}& Considered & Considered & No\\
\textbf{Content categories} & Considered & No & No\\
\textbf{Content density} & Calculated & No & No\\
\bottomrule
\end{tabular}
\label{scheme comparison}
\end{table}

\begin{figure}[!t]
	\centering
	\includegraphics[width=3.5in]{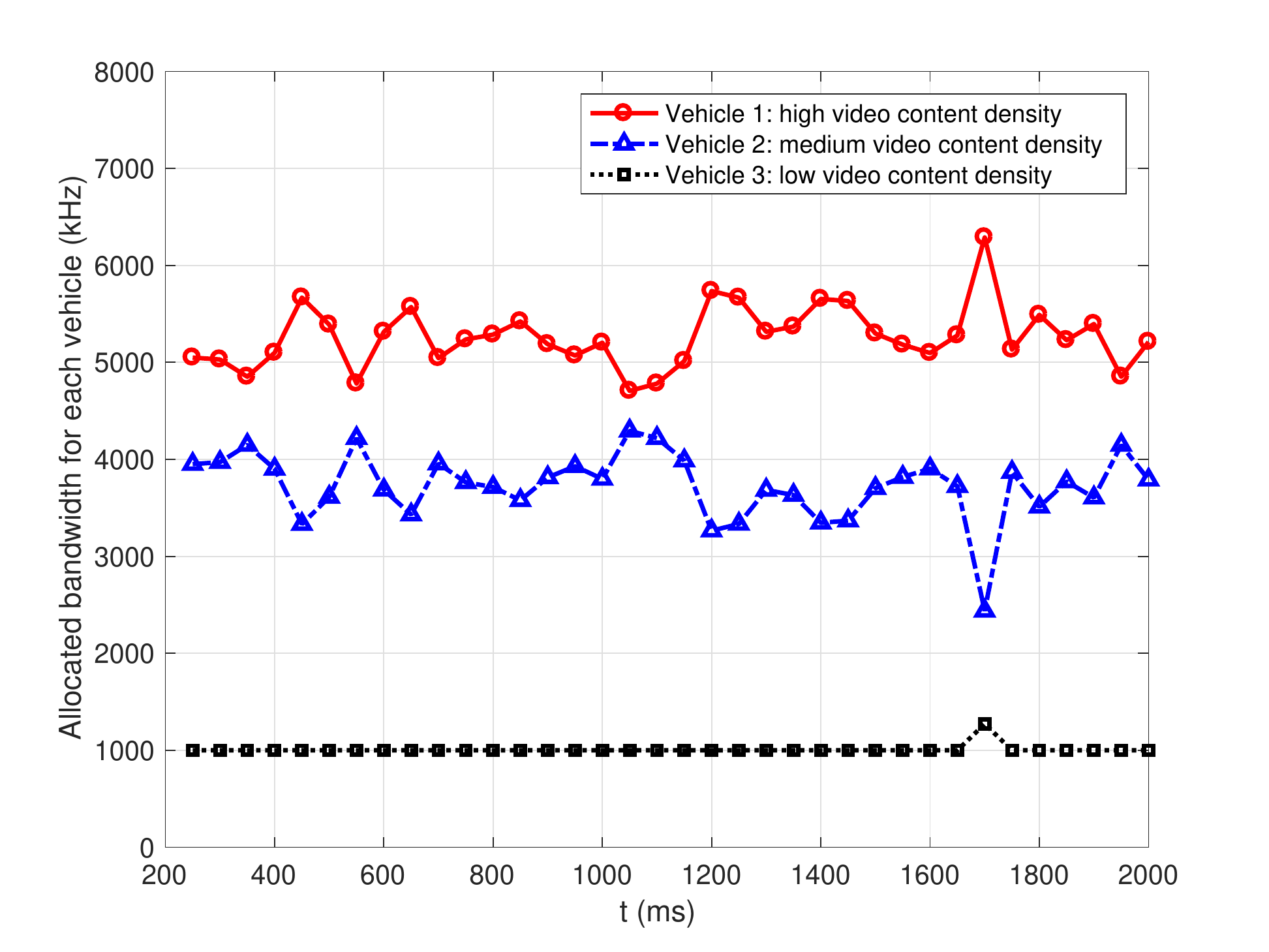}
	\caption{QoC-based bandwidth allocation results with time-varying channel.}
	\label{time varying allocation}
\end{figure}

Figure {\ref{time varying allocation}} shows the results of QoC-based bandwidth allocation with time-varying channel state based on Eq. (\ref{eq5}), the all transmission time is $T=2\rm s$, and the total bandwidth is $B_{\rm{total}}=10\rm{MHz}$. Clearly, the proposed scheme tends to provide higher bandwidth for the video with higher detection objectives density to ensure the quality of video content. Moreover, since channel is not constant with vehicle mobility, the transmission rate from vehicle to the edge server changes over time, so the result of resource allocation changes during a video transmission period. Moreover, with rapid channel changes, frequent updates of resource allocation based on proposed scheme ($L=(T-\Delta t)/t_e=36$ , in Fig.{\ref{time varying allocation}} ), so the proposed scheme can adapt to the dynamic environment.

\begin{figure}[!t]
	
	\centering
	\includegraphics[width=3.5in]{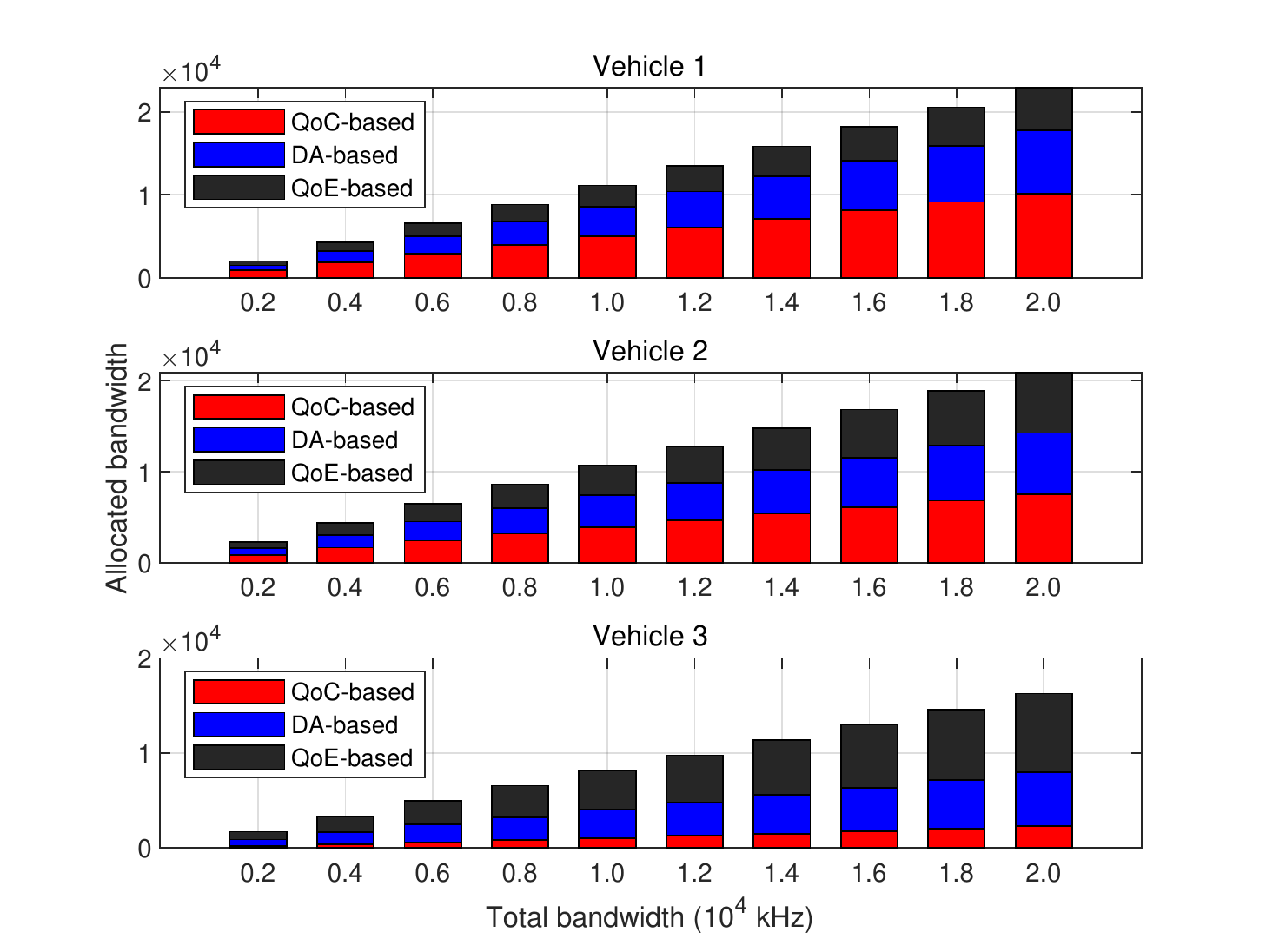}
	\caption{Bandwidth allocation of the 3 vehicles with different schemes.}
	\label{Bandwidth allocation}
\end{figure}

Figure {\ref{Bandwidth allocation}} shows the bandwidth allocation results with different schemes. With the proposed scheme, vehicle 1 is allocated more bandwidth since its collected video contains more content information (given in Table \ref{video information}). Since the QoE based scheme is irrelevant to video content information, its allocation results is quite different. With the DA driven scheme, video with low content density like video 3 is also allocated some bandwidth for the pursuit of uniform detection accuracy. The results reveal that more informative videos have advantage on resource allocation under proposed scheme.

\begin{figure}[!t]
	\centering
	\includegraphics[width=3.5in]{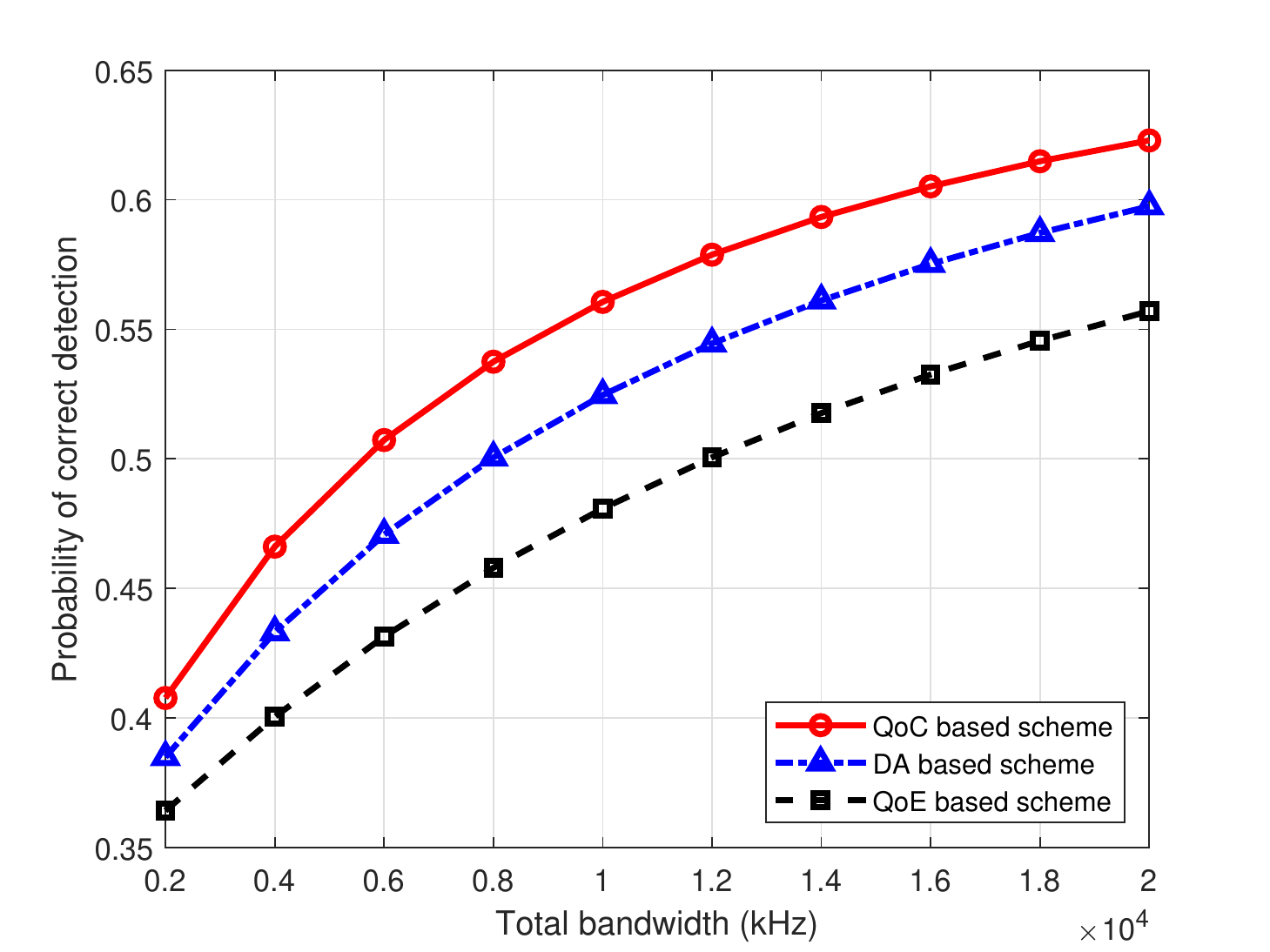}
	\caption{Overall probability of correct detection with different schemes.}
	\label{Overall probability}
\end{figure}

The overall detection accuracy performance  with different schemes are shown in Fig.{\ref{Overall probability}}. In the simulation, the detection accuracy is equal to true positive detection probability. For all schemes, the higher the total bandwidth, the higher the probability of correct detection, since higher bandwidth ensure better video quality. Due to the limitation of the detection system, the probability curves become flat out when the total bandwidth resource is sufficient. According to the results, the proposed scheme achieves better performance on detection accuracy due to its effective bandwidth allocation results.

\begin{figure}[!t]
	\centering
	\includegraphics[width=3.5in]{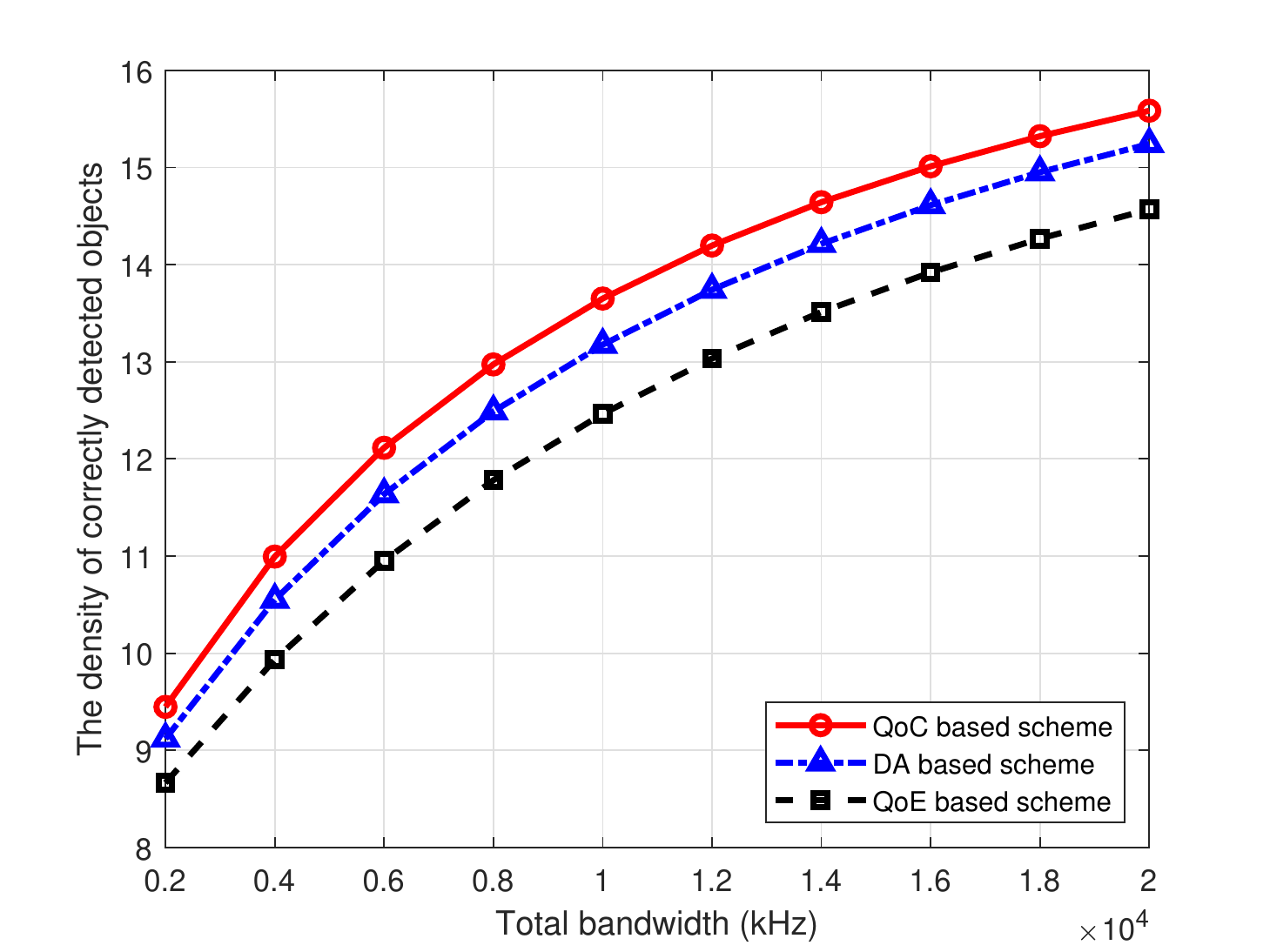}
	\caption{Overall density of correctly detected objects with different schemes.}
	\label{The overall density}
\end{figure}

The overall density of correctly detected objects with different schemes are shown in Fig.{\ref{The overall density}}. We use the overall density of object to represent the amount of content information, defined in Eq.\ref{eq3}. Obviously, the proposed scheme leads to higher density of correctly detected objects. That is, the proposed scheme ensure that more objects can be detected correctly at the edge server again after video transmission. The results illustrate that efficient resource allocation helps to retain more content information of videos in transmission, so more useful content information can be received at the server.

\section{Conclusion}
In this paper, we proposed a QoC based bandwidth allocation scheme for video content understanding in vehicular network. Different from the traditional wireless resource allocation schemes such as QoS or QoE based schemes, our proposed scheme devotes to improve the overall content quality of transmitted videos. The Faster R-CNN object detection system is used for vehicles and its detection accuracy in terms of QP is given. Then, we derived the quality assessment of video content. The proposed bandwidth allocation scheme to maximize the overall correct detected probability can be transformed into a convex optimization problem. Simulation results demonstrate the effectiveness of our proposed scheme. For future works, combining wireless video transmission resource management and computer vision techniques is interesting and crucial for automatic driving applications.

\appendices
\section{Proof of the convexity of P2}
For the definition of channel model in Section III, near perfect CSI was known, and small-fading defined in Eq.(\ref{eq5}) is constant in duration $t_e$. In time $t\in (\Delta t, \Delta t+t_e)$, the channel gain $h_m$ is constant and not time-varying. Hence, firstly, according to Eq.(\ref{eq4}), transmitted rate $R^T_m(B_m) = B_{m} \log _{2}\left(1+\frac{S_{m} h_{m}}{\sigma^{2}}\right)$ is a logarithmic function, which is concave and increasing, with the independent variables $B_m$. Secondly, according to Eq.(\ref{eq2}), $Q_m(\cdot)$ is convex and decreasing, so the function $Q_m(R^T_m(B_m))$ is convex and decreasing by the composition rule. Thirdly, according to  Eq.(\ref{eq1}), $P_n(\cdot)$ is a concave and decreasing function, so composite function $P_n(Q_m(R^T_m(B_m)))$ is concave. Furthermore, as shown in Table \ref{model parameters}, $\alpha_n<0$, $\beta_n>1$ and $b_m<0$, then $\frac{\partial P_{n}\left(Q_{m}\left(R_{m}\left(B_{m}\right)\right)\right)}{\partial B_{m}}>0$. Lastly, due to $\delta_n$ and $I_{m,n}$ are known constants, $M$ and $N$ are finite positive integers, the objective function of P2 is non-negative sums of concave and increasing functions, based on the rules of composite functions.

Therefore, the optimization problem in P2 is convex, since the objective function is concave and the feasible set determined by all the constraints in P2 (C1$\sim$C4) is convex.

\end{document}